\documentclass[preprint2]{aastex}

\usepackage{epsfig,amssymb}
\usepackage[dvips]{color}
\usepackage{graphicx}

\shorttitle{Inefficient enrichment of molecular clouds}
\shortauthors{Ntormousi \& Burkert}

\begin{document}


\title{Delayed metal recycling in galaxies: 
the inefficiency of cold gas enrichment in colliding supershell simulations}


\author{Evangelia Ntormousi \altaffilmark{1,2} and Andreas Burkert \altaffilmark{1,2,3}} 


\altaffiltext{1}{University Observatory Munich, Scheinerstr. 1, D-81679 Germany}
\altaffiltext{2}{Max Planck Institute for Extraterrestrial Physics, Giessenbachstr.
85748 Garching Germany}
\altaffiltext{3}{Excellence Cluster Universe, Boltzmannstr. 2
D-85748 Garching Germany}


\begin{abstract}

The fate of metals ejected by young OB associations into the Interstellar Medium (ISM) is investigated numerically.
In particular, we study the enrichment of the cold gas phase, which is the material that forms molecular clouds. 
Following previous work, the expansion and collision of two 
supershells in a diffuse ISM is simulated, in this case also introducing an advected quantity which represents the metals
expelled by the young stars. We adopt the simplest possible approach, not differentiating between metals
coming from stellar winds and those coming from supernovae.
Even though the hot, diffuse phase of the ISM receives a significant amount of metals from the stars,
the cold phase is efficiently shielded, with very little metal enrichment.   
Significant enrichment of the cold ISM will therefore be delayed by at least the cooling time of this hot phase.
No variations in cloud metallicity with distance from the OB association or with direction are found, which means
that the shell collision does little to enhance the metallicity of the cold clumps.
We conclude that the stellar generation that forms out of molecular structures, triggered by shell collisions cannot be significantly enriched.

\end{abstract}




\section{Introduction}

Stellar feedback is a very powerful source of thermal and turbulent energy in the Interstellar Medium (ISM).  
Young massive stars produce ionizing photons, expel large amounts of mass and momentum in winds 
and end their lives in supernova explosions, all processes which shape the matter around them in shells and cavities \citep{Heiles_79, Heiles_84, Ehlerova_05, Churchwell_06}.  
Young OB associations, typically containing between 10 and 100 stars, create 
giant shell-like shocks from the combined feedback of their member stars. 
These supershells are among the largest structures in the Galaxy. 
      
Massive stars are also believed to be an important source of metals for the ISM.  According to our standard 
picture of the chemical evolution of galaxies, material ejected by the stars in an OB association
will take part in forming new stars.
But, how long does it take before the metals released by one generation of stars actually end up 
in molecular clouds?

Molecular clouds collapse gravitationally to give stars very soon after they come 
into existence \citep{Hartmann_2002}, so
any metals they contain are most likely in place when they form.
Simulations of molecular cloud formation are then a natural way of answering 
the question posed above.  
Usually, such simulations fall into two categories:
Galactic scale simulations, which study the gravitational collapse of gas 
caused either by global disk instabilities \citep{KO_2002,KO_2006} or by the self-gravity
of large amounts of gas accumulated by spiral density waves \citep{Dobbs_11b,Dobbs_11}, and local simulations 
of converging atomic flows, which show that atomic gas converts to molecular material
under the combined effect of various fluid instabilities \citep{Heitsch_2005,Heitsch_2006,
Hennebelle_MCs_2007,VS_2007}.       
These approaches could lead to different answers regarding the enrichment of molecular clouds.

In \citet{EN_11}, hereafter Paper I, we presented the results of superbubble collisions in a uniform and in a turbulent diffuse environment.
The energetic feedback from young OB associations, inserted as a time-dependent mass and energy source in the 
numerical simulations, condensed the gas in large spherical shocks.

One indirect conclusion of that work was that the clumps which form from the fragmentation of the shells
seemed to be largely composed by diffuse 
Interstellar Medium (ISM) material, swept-up and condensed by the shocks.
However, even in such a case, some enrichment may still be possible as a result of
the turbulence generated by the shell collision.
To resolve this issue, in this work we investigate the possible metal enrichment of clumps formed 
in such an environment in detail.

There is little previous work studying the fate of the metals ejected by OB associations.
\citet{Tenorio-Tagle_96} considered all the possible mechanisms for mixing between the different
phases in the dense shells of single supernova remnants and in supershells formed by combined explosions.
With dimensional arguments he showed that the mixing should be most efficient in the hot phase of the gas.  
\citet{Spitoni_08} studied the fate of dense clouds formed by the condensation of supershells and  
expelled from the galactic disk in the form of an outflow.  However, they focused more on the landing coordinates
of these clouds when they fall back on the disk.  

In this work for the first time we study the
efficiency of the metal enrichment of the cold phase formed at the edges of supershells numerically.  
We adopt a very simple approach, in which the OB associations inject a constant amount of metals with time.  

Our methods is presented in Section \ref{nm}, the results are discussed in Section \ref{res} and
we comment on their implications in Section \ref{concl}.
 
   
\section{Numerical Method}\label{nm}

A two-dimensional, high-resolution simulation very similar to those in Paper I 
is performed with the hydrodynamical code RAMSES \citep{Teyssier_2002}.
The setup consists of two young OB associations, placed at a certain distance from each other in a 
warm diffuse background.  Their feedback creates two expanding shells that sweep up 
the surrounding gas.  These cold and dense shells
eventually collide in the middle of the computational domain.    

The time-dependent wind and supernova feedback from these stellar associations
is implemented as a source of energy and mass in the code.
An OB association in this simulation comprises 20 "average" stars.  
Each of these stars represents a fraction of an entire population. 
For simplicity, all stars are placed in a circular region of 5 pc radius.  The associations are assumed to form
simultaneously and all stars within each of them are also assumed to have formed at the same time.
The wind and supernova data were taken from population synthesis models created by \citet{Voss_2009}.
For more details on the wind implementation we refer the reader to Paper I.
Cooling and heating processes appropriate for the ISM are also included, according to \citet{Dalgarno_1972},
\citet{Wolfire_1995} and \citet{Sutherland_1993}.  The simulation does not include gravity.

An important increase in efficiency in comparison to previous simulations is achieved with 
the use of Adaptive Mesh Refinement (AMR).  Given the nature of the problem under study, the most adequate
refinement policy is to trigger the division of a cell when the difference in the gradients of pressure and density
exceeds a certain threshold.  We have used a threshold 
equal to 1\% of the gradient of the density and the pressure in this work, but experimenting with the value of the threshold gave no significant differences in the grid structure.  Thresholds higher than 10\%, however, failed to capture the shock structure properly. 

\begin{figure*}
    \includegraphics[width=0.5\linewidth]{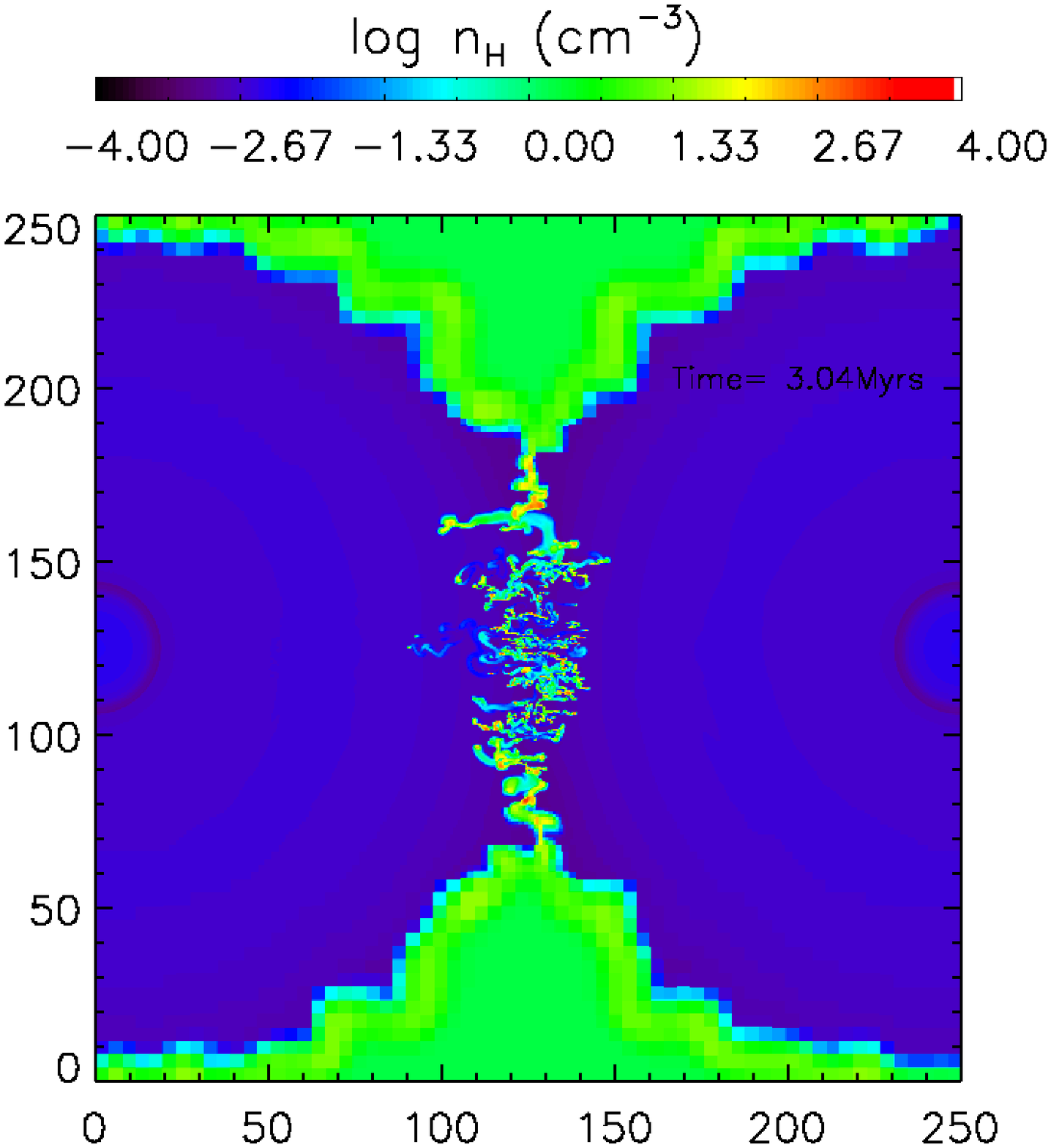}  
    \includegraphics[width=0.5\linewidth]{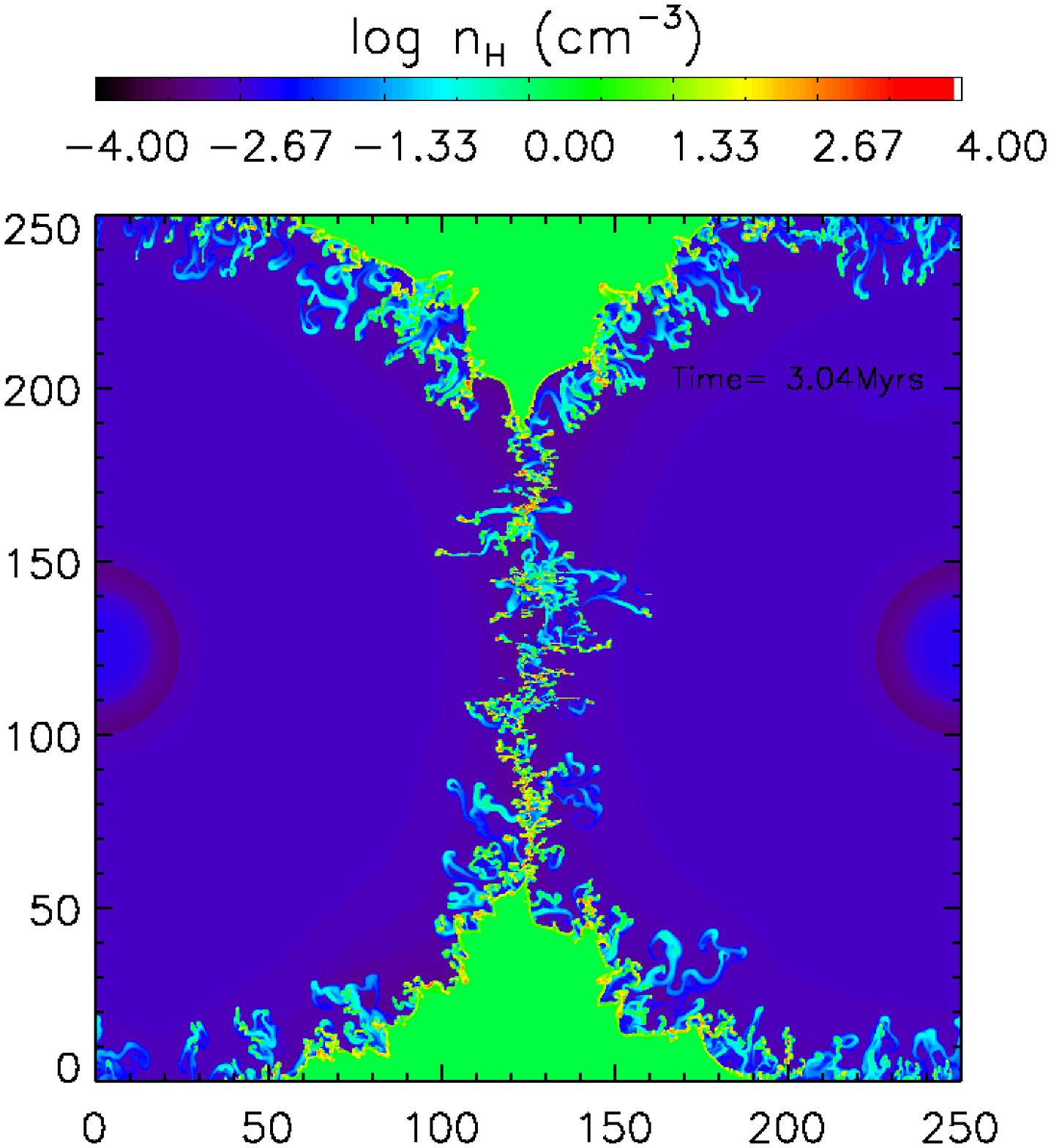} 
    \includegraphics[width=0.5\linewidth]{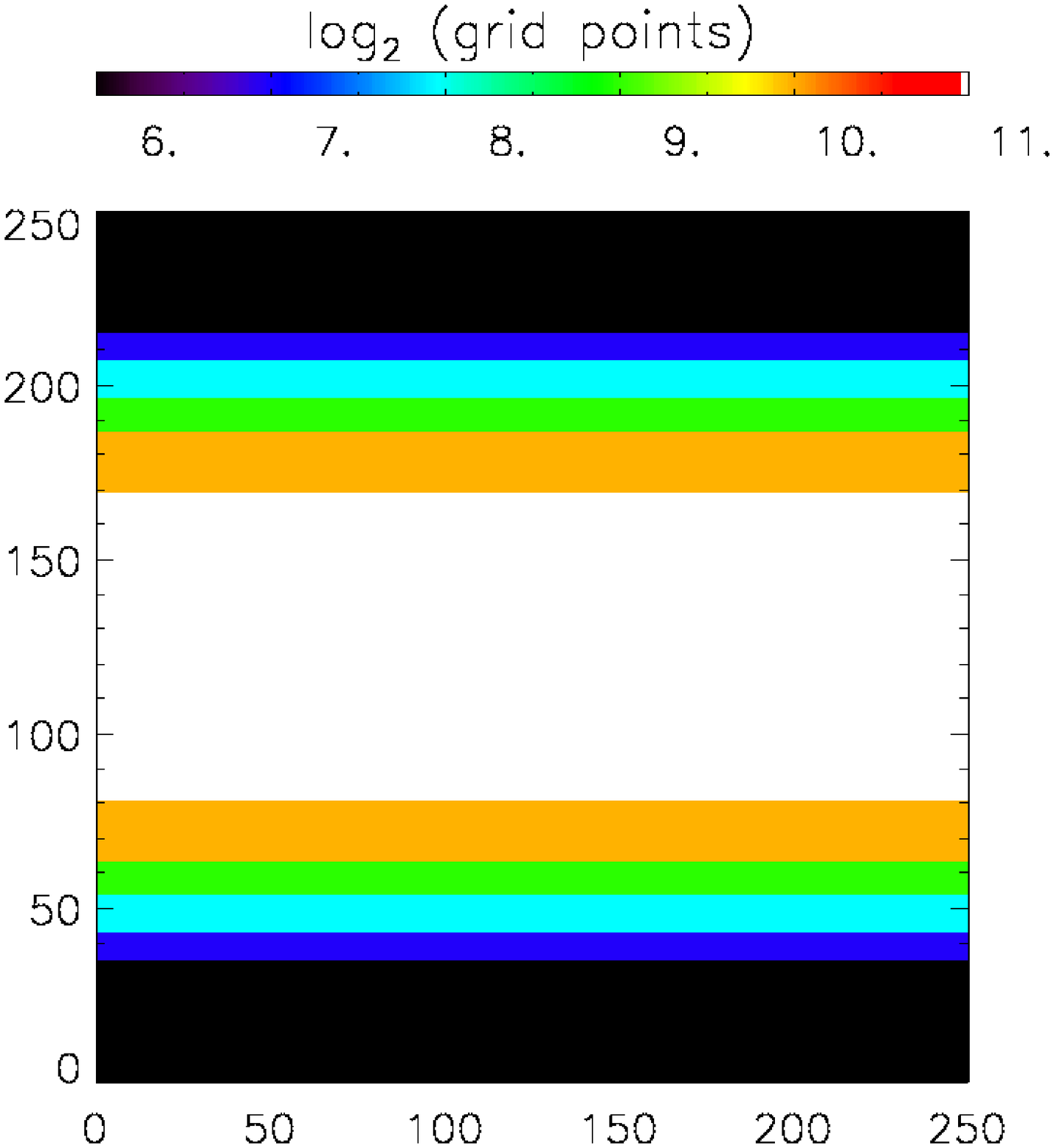}  
    \includegraphics[width=0.5\linewidth]{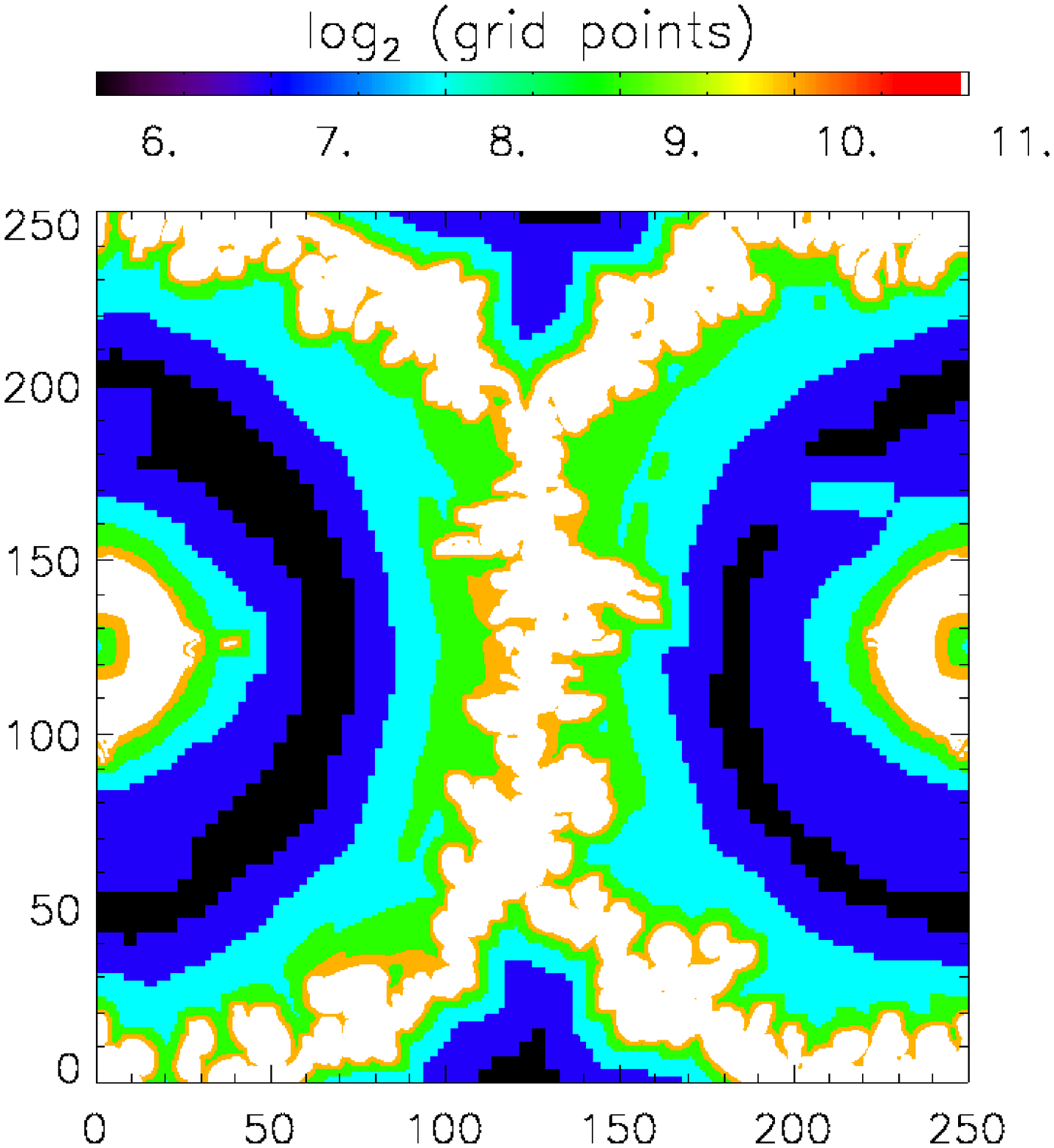} 
    \caption{Snapshots of two runs using different refinement techniques, taken at the same timestep, about 3 Myrs after star formation.  
    On the left, nested grid and  on the right gradient-based refined grid. 
    The plots on the top row show the logarithm of hydrogen number density and the plots on the bottom row show the corresponding grid structure.
    The axes coordinates are in parsecs.}
   \label{amr_comparison}
\end{figure*}

The supershells in this setup are expanding in a uniform diffuse background, which means that any perturbations that are expected
to seed the fluid instabilities must arise at the grid level.
In order to be able to compare this simulation, which uses AMR, to our
simulations from Paper I, which were done with a uniform grid, we must seed the perturbations
at the smallest grid level and at the same physical scale.  
For this reason, the simulation is initiated with a nested grid configuration,
where the highest resolution region is located at the center of the simulation box.
Once the first seeds of the perturbation start to grow, we switch to the adaptive refinement
policy described above.  Figure \ref{amr_comparison} shows the grid structure for a nested 
and for an adaptive grid refinement policy.  The top panel of the Figure shows the logarithm of the hydrogen number density 
and the bottom panel shows the level of refinement in powers of two.  For example, a level of refinement equal to 6
in this notation means that, if the entire domain were simulated at this resolution, it would contain $2^6$ cells.

Comparison between simulations of this setup with this AMR approach and uniform grid simulations
have shown no difference in the amount of cold gas formed in the simulation, the position of the
shocks and the sizes or velocity dispersions of the formed clumps.  Small differences in the
shock morphology are, of course, always present, due to the very nonlinear nature of these phenomena. 

For this particular simulation we use a box of 250 pc physical size, at an effective resolution of 2048$^2$
(a maximum level of refinement equal to 11 according to the notation described above).  This
allows us to resolve the Field length of the warm ISM \citep{Koyama_2002}, without actually implementing numerical thermal conduction.
Higher resolution simulations would require such implementation (see also discussion in Paper I).
The choice of a smaller box with respect to Paper I is both more physical, in terms of the average distance between OB associations in the Galaxy
and it also yields a smaller computational volume for the same physical resolution, thus significantly reducing the computational cost of the simulation.

We stop the calculation when the turbulence in the collision area starts expanding towards the inflow 
boundaries. In this particular case this happens 4.36 Mys after star formation in the OB associations.

The aim of this work is to follow the advection of metals from the OB associations into the cold gas formed at the shock wake.  This is done by means of a passive advected quantity, representing the metals from the stars.
A constant amount of metals, equal to 10$^{-3}$ metal particles/cm$^3$ is added at the wind region at each coarse timestep.
This value is totally arbitrary, so it can be scaled to represent different environments.

The amount of metals introduced by the OB associations is assumed here to have a negligible effect on the amount of cold gas formed.
In principle, though, extra metals could affect our results due to their contribution to the cooling,
which would also change the regime where the gas becomes thermally unstable.
In our calculations the most important coolant of the gas is line emission from carbon and oxygen. 
A decrease in the abundance of these elements 
would cause the area where we can have phase formation due to the Thermal Instability to shrink, and  
an enrichment would enlarge the Thermal Instability regime  \citep{Wolfire_1995}.  

Support for the approximation we are making comes from \citet{Walch_11}, who studied the effect of metallicity on the formation of cold gas from Thermal Instability in 
simulations of turbulence.  They found that, for driven turbulence (which is the case in our models), the total amount of cold
gas in the simulation is not significantly affected by changes in the metallicity.  This means that, as long as the metallicity of the gas we are simulating
is high enough to capture the Thermal Instability regime of the ISM, we are not making significant errors in the total amount of cold gas in the domain by ignoring the enrichment from the OB stars in the cooling function. 

\section{Delayed metal recycling}
\label{res} 

\begin{figure*}
  \includegraphics[width=0.5\linewidth]{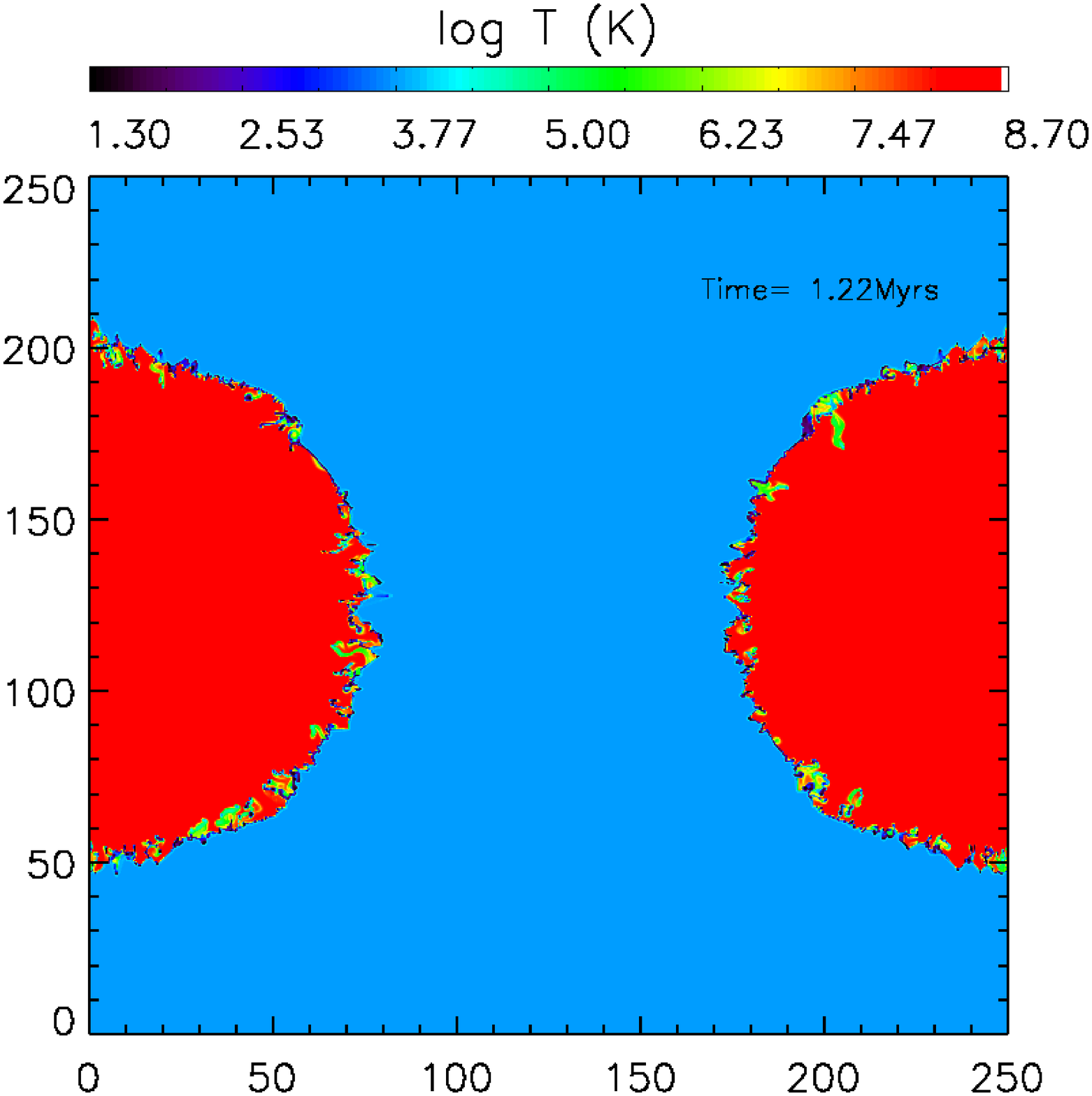}
  \includegraphics[width=0.5\linewidth]{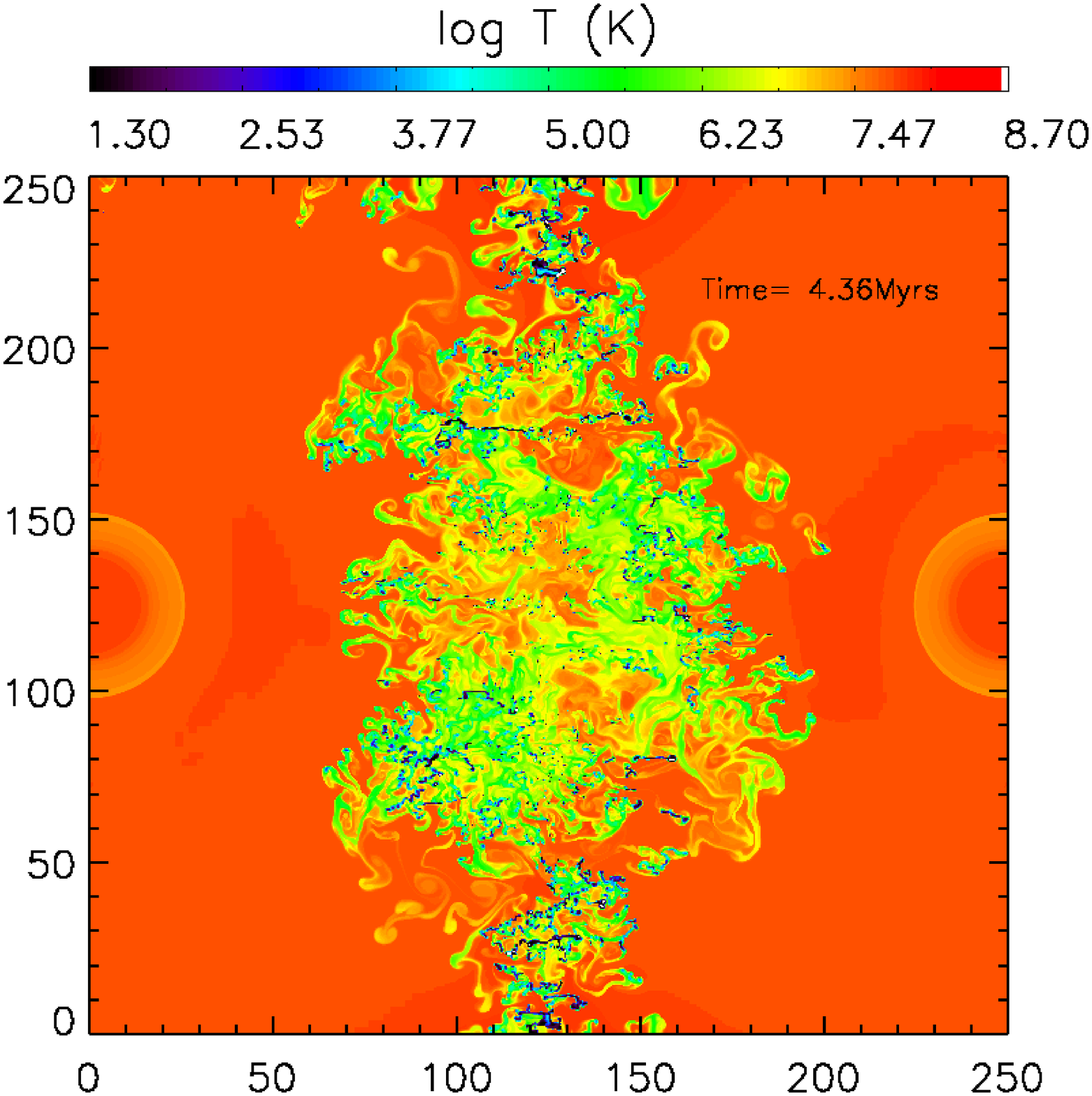}
  \includegraphics[width=0.5\linewidth]{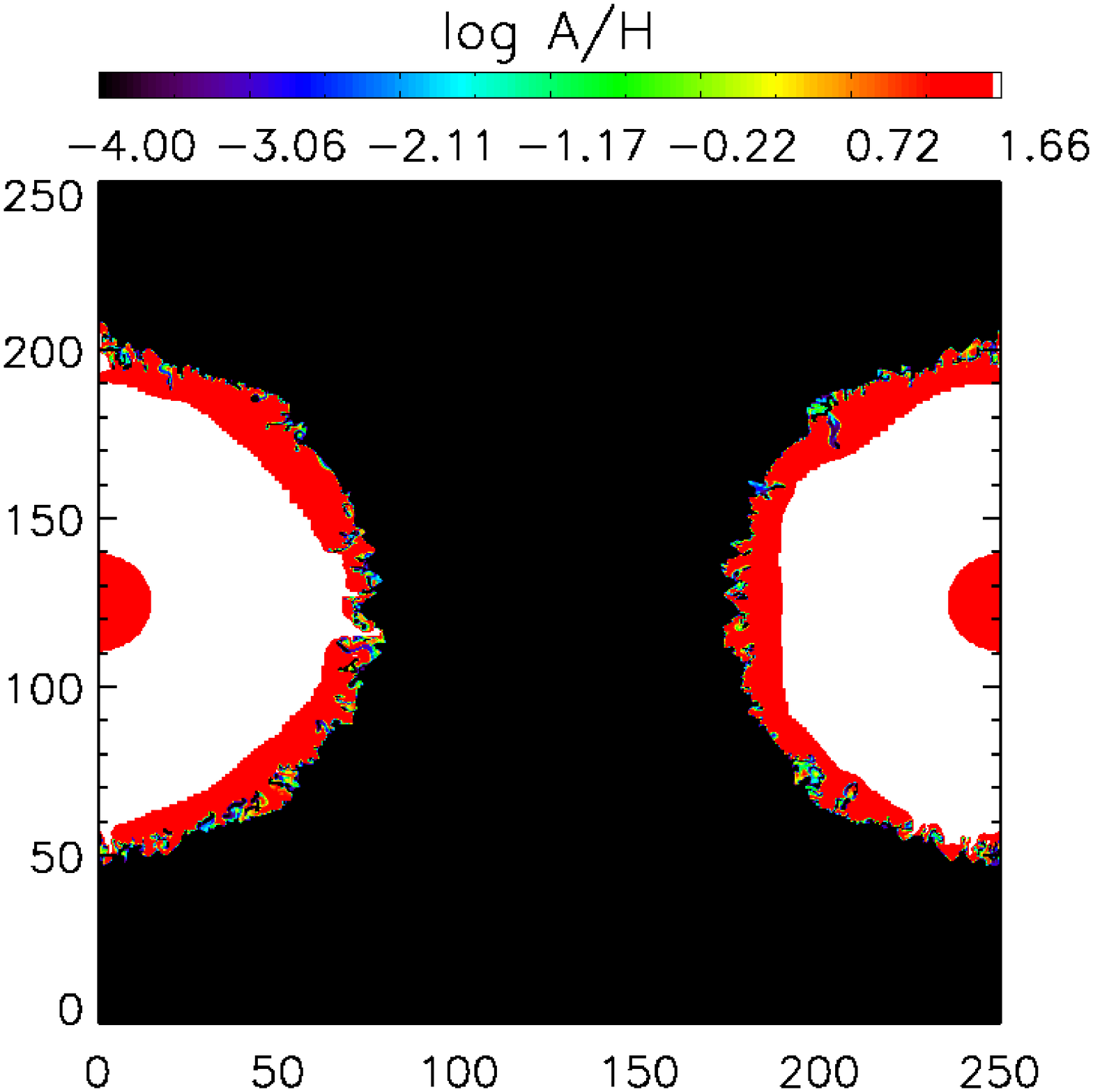}
  \includegraphics[width=0.5\linewidth]{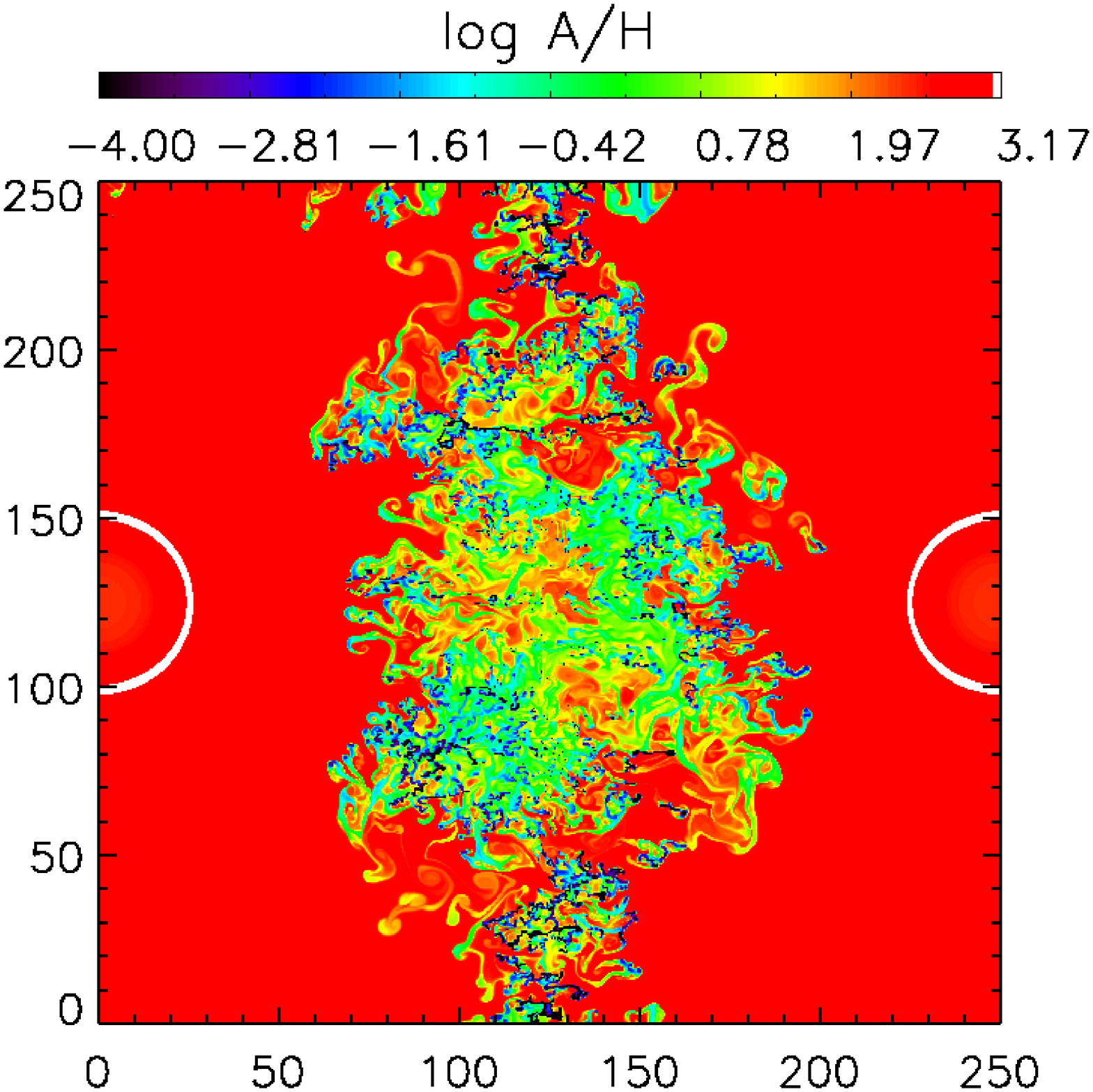}
  \caption{Logarithm of temperature (top) and logarithm of relative metal content (bottom) for two snapshots.  
   On the left, 1.22 Myrs and on the right, 4.46 Myrs  after star formation took place in the OB associations.}
   \label{snapshot_den_tem}
\end{figure*}

Figure \ref{snapshot_den_tem} shows snapshots of the simulation before and after the shell collision.
The top panels of this Figure are contours of the logarithm of the gas temperature and the bottom
panels show the logarithm of the ratio of newly ejected metals to hydrogen atoms in the cell.

The general picture of the simulation is the same as in Paper I.
The spherical shocks created by the stellar feedback are unstable to the Vishniac instability \citep{Vishniac_1983,Vishniac_1994} as small-scale wind fluctuations create ripples on their surface.
The result of the gas condensation at the peaks of these ripples is to trigger the Thermal Instability \citep{Field_1965, Burkert_Lin_2000},
which creates cold and dense clumps at the shock wake.  The shear on the shell surface, also caused by the Vishniac Instability, gives rise to
characteristic Kelvin-Helmholtz eddies, 
thus contributing to the dynamics of the newly-formed cold clumps (right panel of Figure \ref{snapshot_den_tem}).  

When the shells collide, the combination of the large-scale shear by the collision and the 
small-scale structure already carried by the shells gives rise to a turbulent region at the collision interface
which contains a mixture of warm and cold gas (right panel of Figure \ref{snapshot_den_tem}).
Turbulence is a very efficient mixing mechanism, so we expect an enhancement in metallicity
of the warm gas after the shell collision. 

\begin{figure*}
  \includegraphics[width=0.5\linewidth]{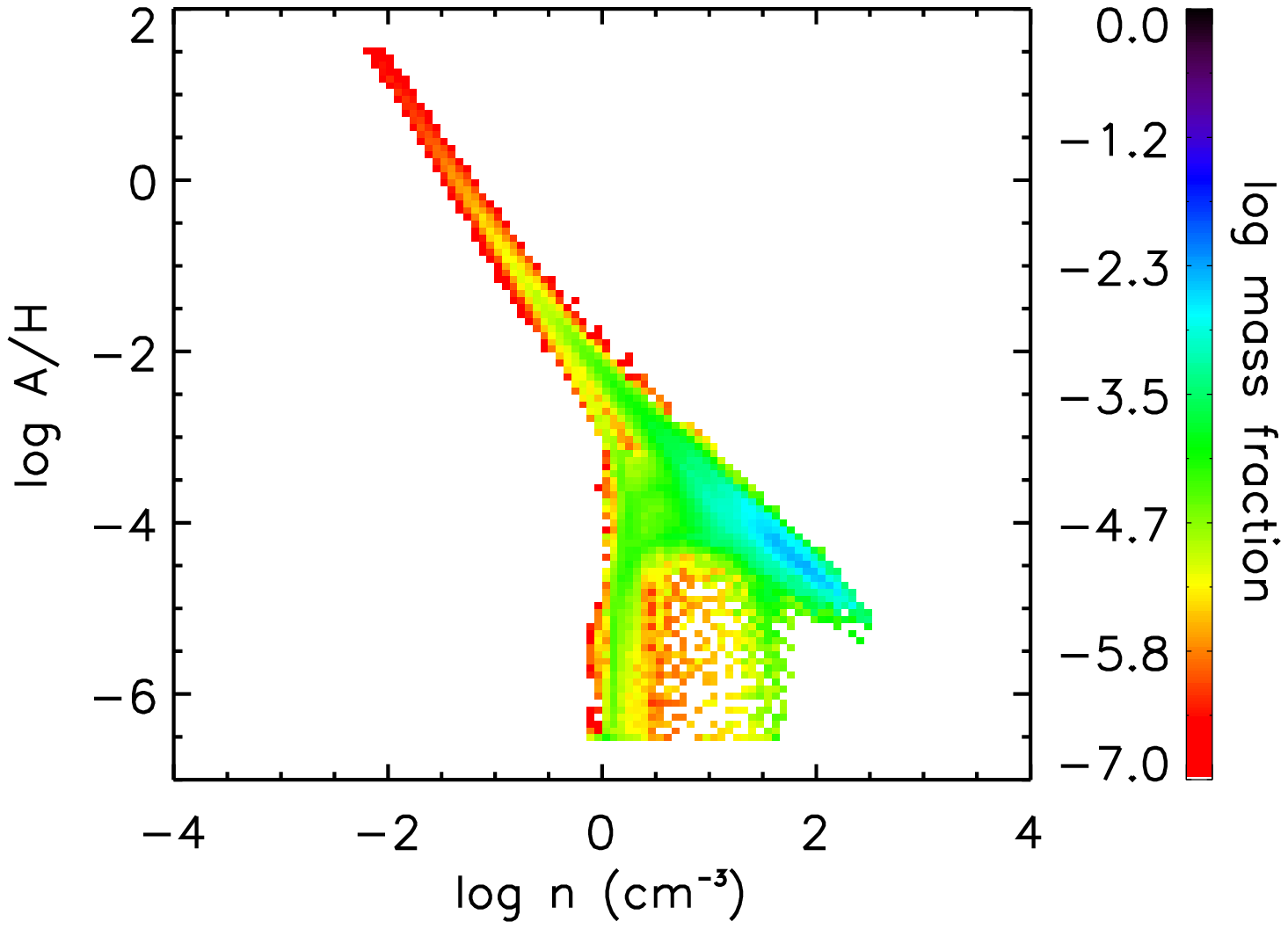}
  \includegraphics[width=0.5\linewidth]{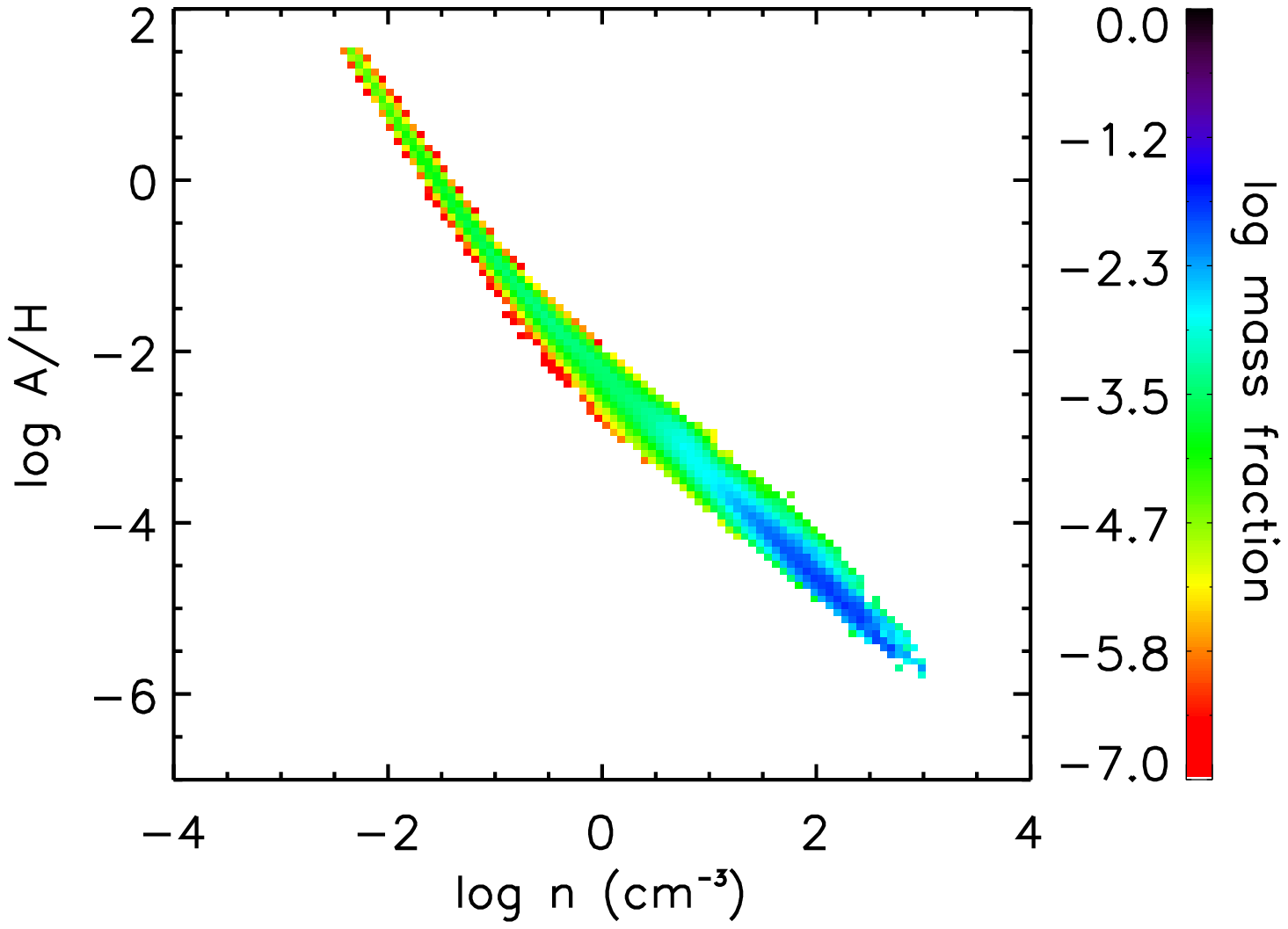}
  \caption{Mass fractions in density-metallicity bins at two snapshots, 1.22 Myrs (left) and 4.36 Myrs (right)
 after star formation.}
  \label{metal_histograms}
\end{figure*}

In Figure \ref{snapshot_den_tem}, we can indeed see that the warm gas
has enhanced metal content. The same can be shown more clearly by plotting the mass fraction of the 
gas in the computational domain in density-metallicity bins.  
In Figure \ref{metal_histograms}, showing such plots for two snapshots of the simulation,  we can see that 
the dense gas dominates the mass of the gas in the computational domain.  
At the same time we see that it never reaches relative enrichment of more than 10$^{-4}$.
For comparison, we note that, were the metals ejected by the stars to be instantaneously and homogeneously
mixed in the diffuse gas phase, the relative enrichment would be $10^{-2}$ and if all the metals from
the stars ended up in the cold phase, the relative enrichment of that phase would be of about $5\cdot10^{-2}$.

Throughout the simulation practically all the metals injected by the OB associations stay in the hot wind,
despite the fact that most of the mass is in the cold gas component.
At late times a small fraction of the metals (1-5\%) mixes into the slightly denser, warm gas ($n_H\simeq 10^{-1}$, $T\simeq10^5$)
due to the shell collision that causes turbulent mixing.
 
\subsection{Clump metallicities}

A clump is identified as a collection of adjacent cells with densities above 50 cm$^{-3}$ and temperatures lower than 100 K.
By this definition, Figure \ref{metal_histograms} already indicates that the clumps do not contain significant amounts
of material from the OB associations.

\begin{figure*}
  \includegraphics[width=0.5\linewidth]{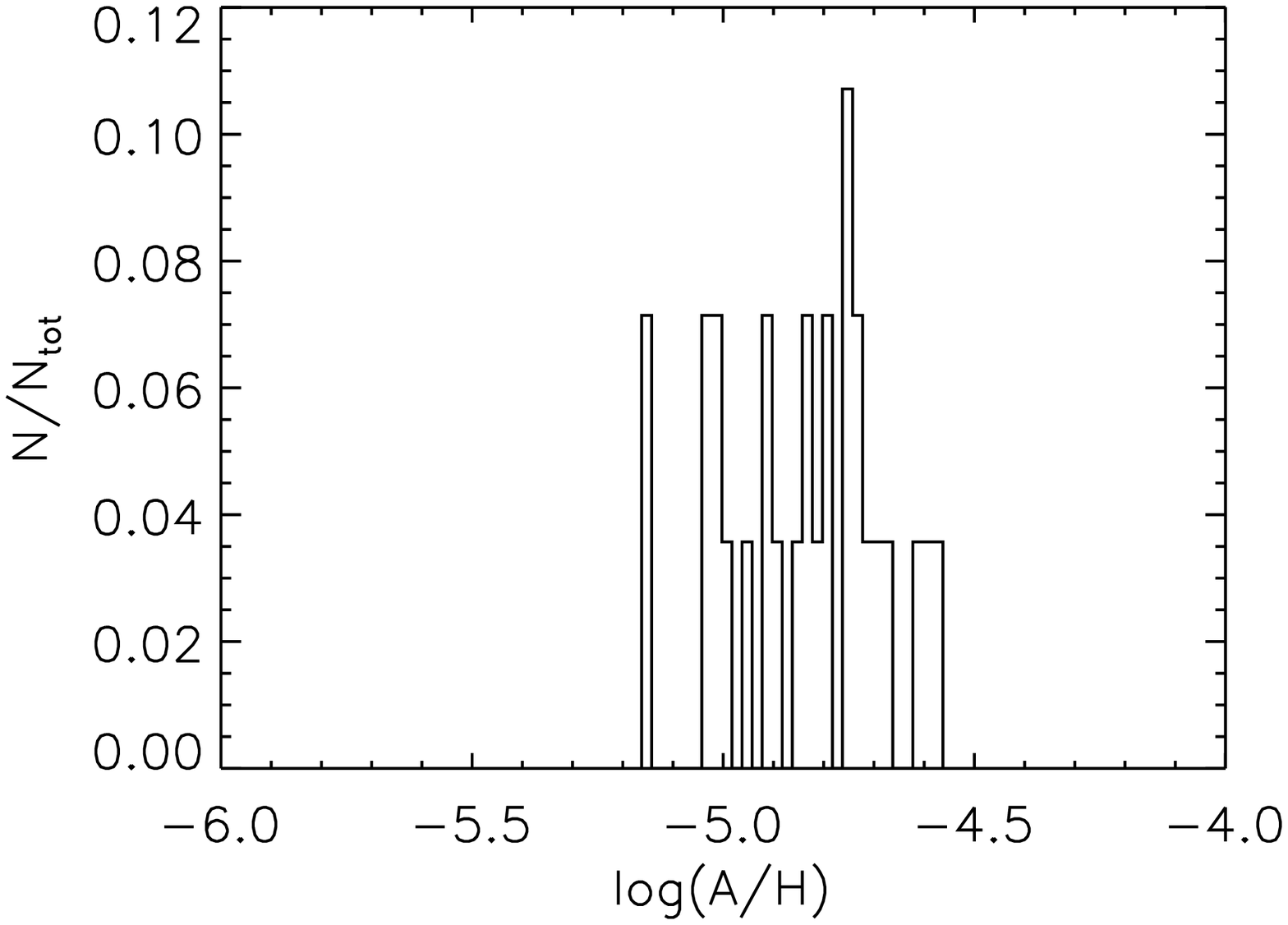}
  \includegraphics[width=0.5\linewidth]{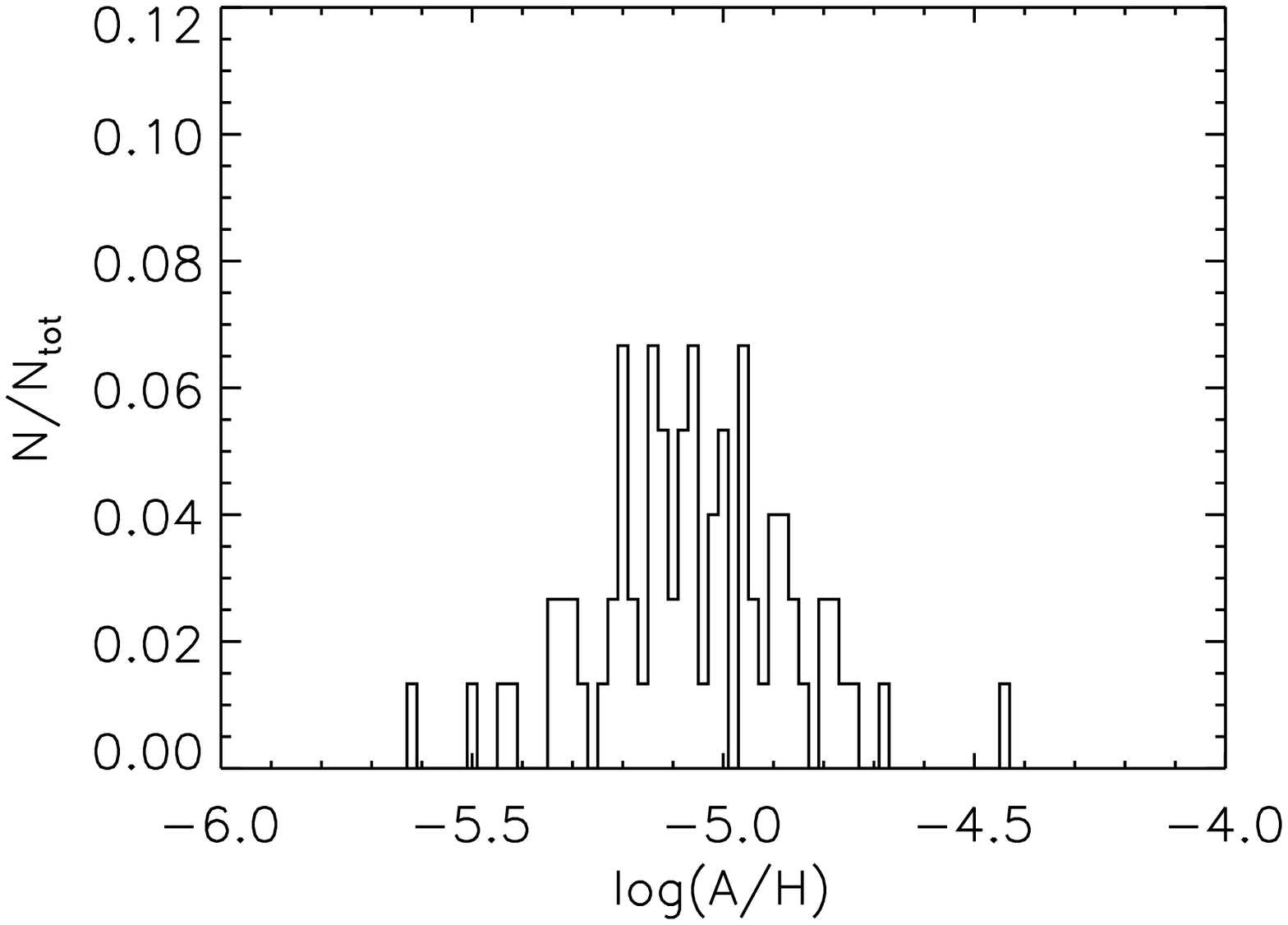}
  \caption{Distributions of the metal content of the clumps over their hydrogen number density.
  The data are from snapshots 1.22 Myrs (left) and 4.36 Myrs (right)
  after star formation.}
  \label{core_histograms}
\end{figure*}

To look at the clump metallicities in more detail, we plot their metallicity distributions
in Figure \ref{core_histograms}.  The plots 
show distributions of the mean metallicity over the mean hydrogen number density of the clumps,  
on the left-hand side for a snapshot at 1.22 Myrs and on the right-hand side
for the final snapshot, at 4.36 Myrs.

Even though the numbers can be rescaled to mean 
different absolute metal content in the clumps, the important fact here is that the cold phase will always receive at least two orders of magnitude 
less metals than the diffuse warm phase.

Even though the metal injection from the OB associations does not stop during 
the simulation time, the metal content of the clumps does not seem to increase significantly.
The spread of the distribution of the relative metal content of the clumps seems to increase with time.
As the system evolves, new clumps are formed at relatively lower metallicities.  
The little metals they accumulate over time leads to the formation of a peak in the distribution.
However, the maximum value of the distribution does not increase, meaning there
is no significant enrichment.

\begin{figure*}
  \includegraphics[width=0.5\linewidth]{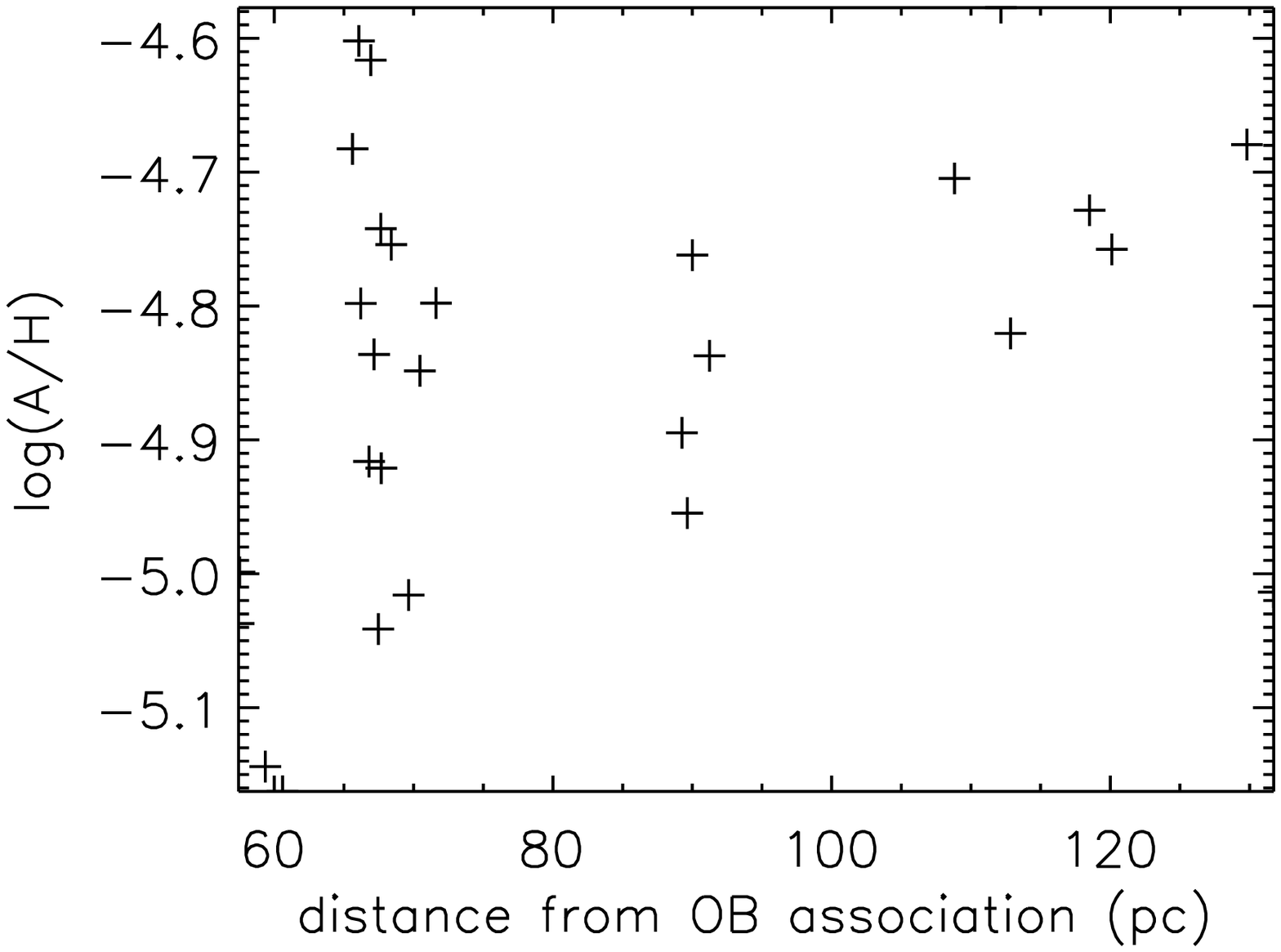}
  \includegraphics[width=0.5\linewidth]{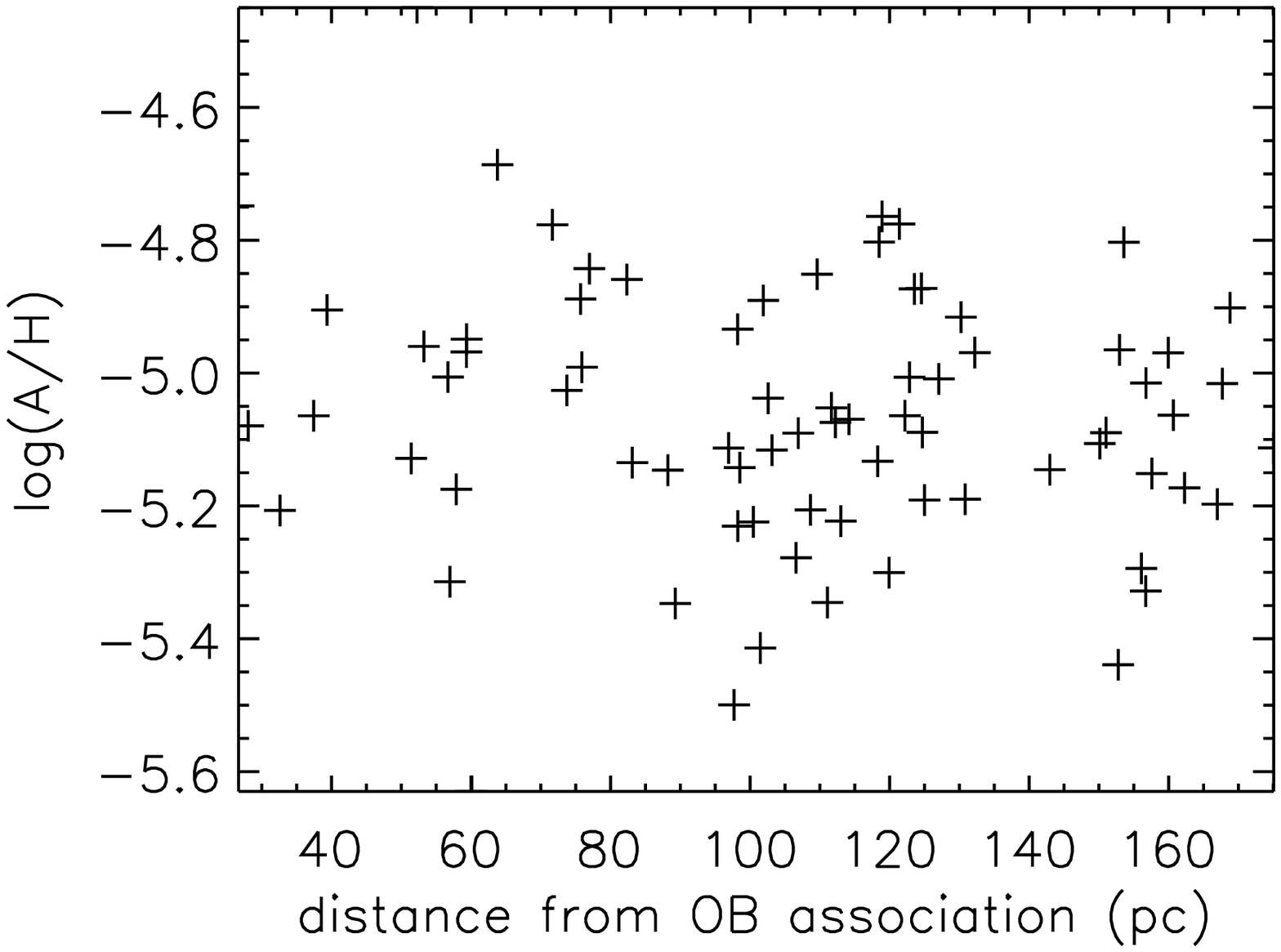}
  \caption{Dependence of the clump metal content on their distance from the closest association.  
  The plots are shown at 1.22 Myrs (left) and 4.36 Myrs (right) since the beginning of the simulation.}
  \label{metals_withrad}
\end{figure*}

\begin{figure*}
  \includegraphics[width=0.5\linewidth]{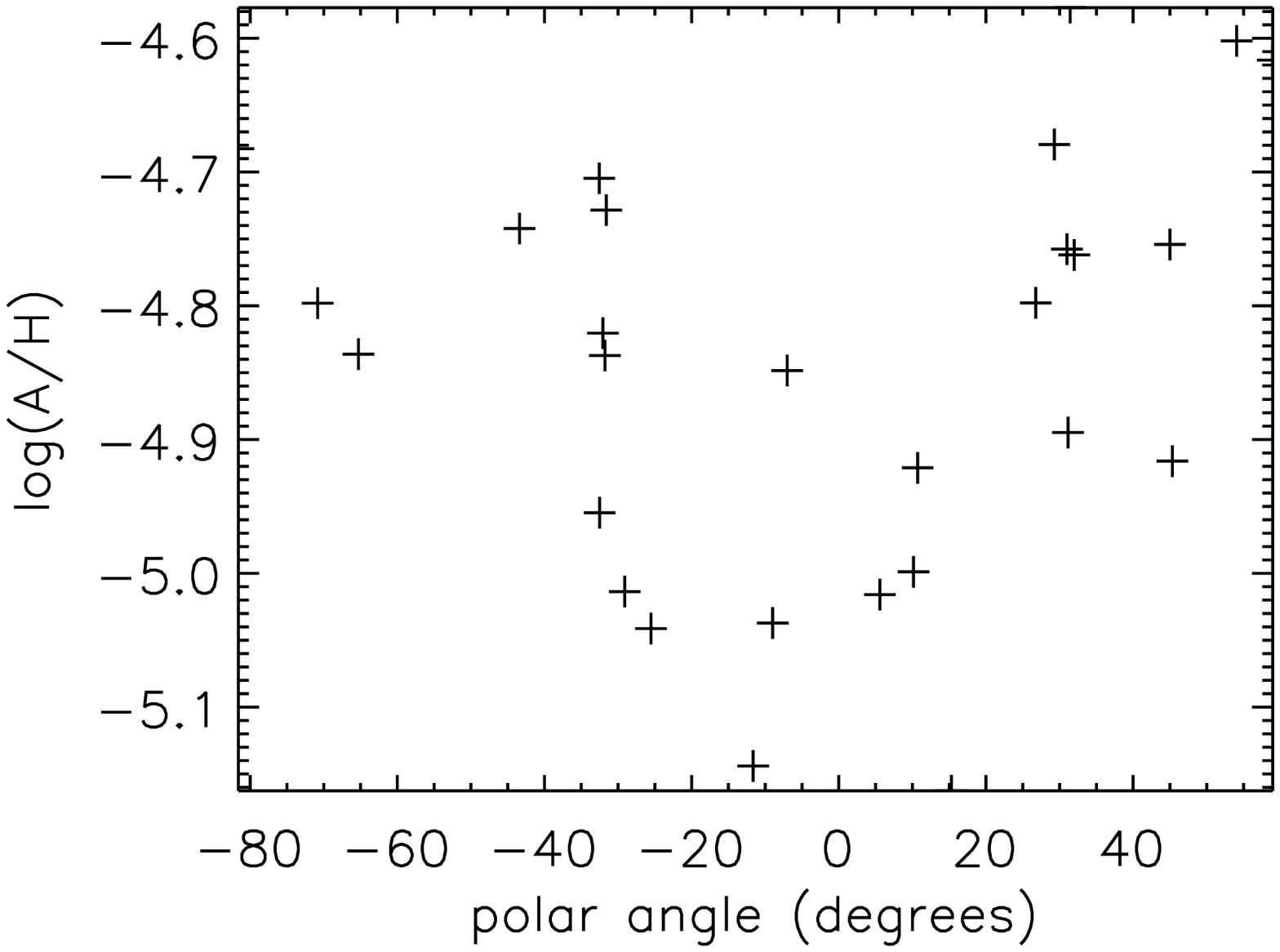}
  \includegraphics[width=0.5\linewidth]{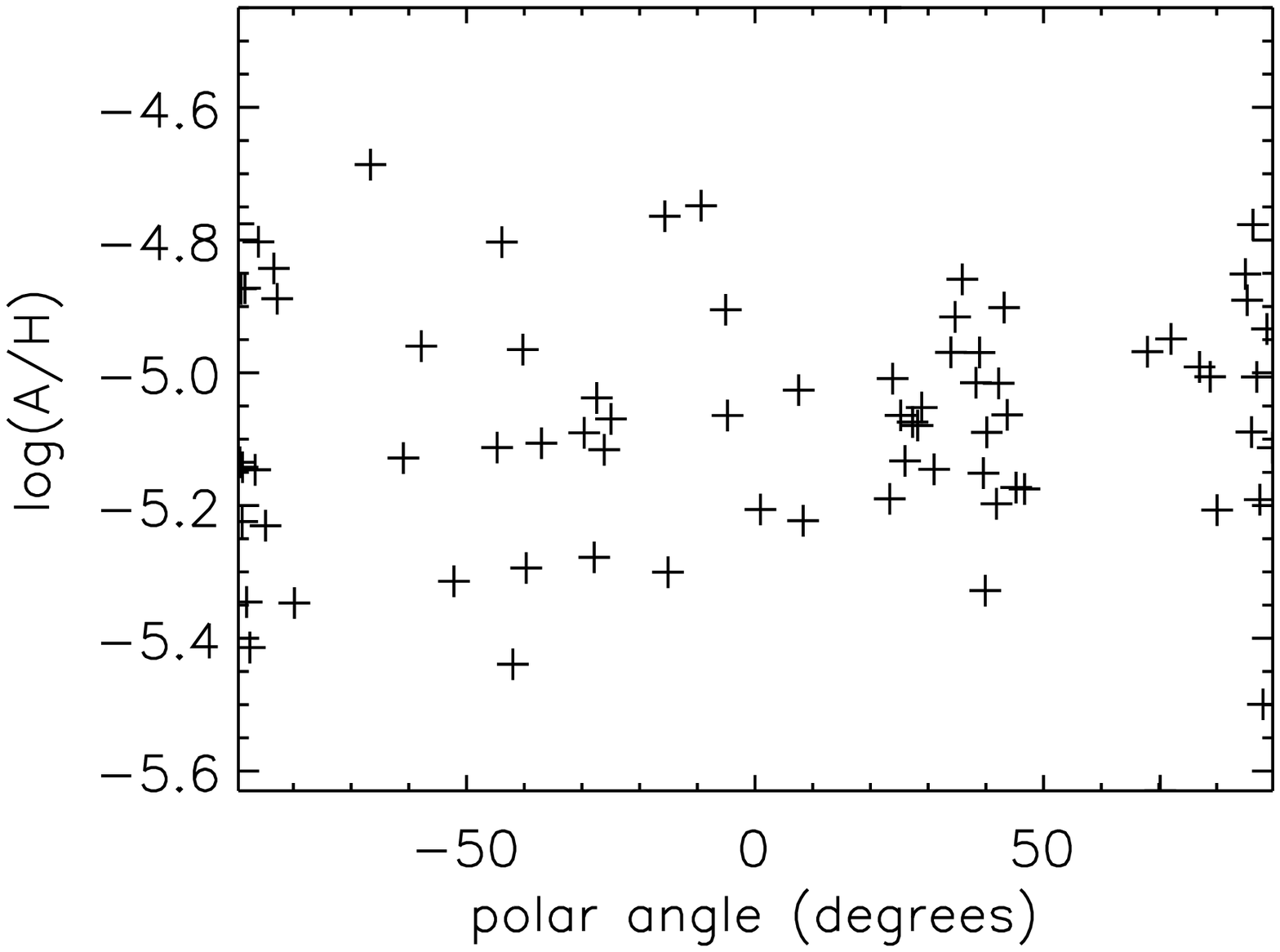}
  \caption{Dependence of the clump metal content on their polar angle 
  calculated with respect to the horizontal line at the center of the computational domain.
  The plots are shown at 1.22 Myrs (left) and 4.36 Myrs (right) since the beginning of the simulation.}
  \label{metals_withphi}
\end{figure*}

Figures \ref{metals_withrad} and 
\ref{metals_withphi} show the mean number density of metals in a clump
as a function of distance from its closest OB association and as a function
of the polar angle with respect to the horizontal line in the middle of the 
domain, respectively. The amount of metals in a clump does not seem to depend on its position
with respect to the OB associations, pointing to a very uniform distribution
of the molecular cloud metallicities around the young associations.  


\section{Conclusions}

\label{concl}

We have presented results of a high-resolution simulation of colliding supershells
created by the feedback from young stars.
The evolution of these shells was followed with ideal hydrodynamics 
and no gravity for about 4.3 Myrs.  In agreement with the results of previous work we observe
the formation of complex cold structure as these shells condense, fragment
and collide.

In this simulation we have used an advected quantity, inserted
in the region of the domain representing the wind, to follow the
metals ejected by the winds and supernova explosions in these OB associations.
In this way, we were able to distinguish between material originating from
the stars and material originating from the diffuse ISM in the 
composition of the cold clumps.

We find that the  
metal enrichment of the clumps is very small throughout the simulation.
The maximum relative metallicity reached by the cold gas in the simulation is 
two orders of magnitude lower than that of the warm ($25000 < T < 10^6$ K) diffuse medium
and negligible compared to that of the hot ($T>10^6$ K) medium.  
The fact that the hot gas receives a significant fraction of the injected metals
implies that, if molecular clouds were to form in this environment, 
enrichment would be delayed by at least the cooling time of this gas.
This effect is even more relevant if we consider that the free-fall time for
each of these dense clumps is about 1 Myr, which means that many of them would 
be collapsing before the end of this simulation, had gravity been considered, 
leaving even less time for enrichment.  We predict that the stars that
are formed in molecular clouds triggered by shell fragmentation
should not be significantly enriched, even taking shell collisions into account. 

A delay in the metal mixing before star formation has important implications for the chemical evolution of galaxies.
The traditional approach for studying the evolution of the metallicity in a galaxy is the closed-box model,
\citep{Searle_72, Tinsley_74}, where mixing is assumed to be instantaneous.  However, 
it has been shown that, at least for the solar neighborhood, better agreement between the
observed metallicities and the models is obtained when this assumption is relaxed.
\citet{Thomas_98} showed that a delay of the order $10^8$ years in the enrichment
of the star-forming gas results in a better fit of the model to local yields.  
\citet{Spitoni_09} came to a similar conclusion, for stellar yields depending on metallicity.

In our simulations the delay in enrichment of the cold gas could be even longer than
the $10^8$ yrs quoted in \citet{Thomas_98}.  The winds from the OB association  
would keep the hot gas at temperatures of $10^7$ K for at least 30 Myrs.  The cooling time
for this gas, given its very low density ($n\simeq10^{-3}$ cm$^{-3}$) is very long 
($t_{cool}\simeq nk_BT / \Lambda\simeq10^{9-10}$ yrs).  This provides further support
for the assumptions of previous 
work regarding delayed metal enrichment of the star-forming phase of the ISM.
 
The metal content of the clumps seems to be independent of their position with respect to the
OB association.  This, in combination with the small spread in cloud metallicities, means that
the next stellar generation, formed by the clumps created in such an environment, would 
be very uniform in its metal content.

Of course, there are many effects that have not been included in this work.
For instance, we have assumed that the metal mass injected by the OB associations
is roughly constant with time and that it is uniformly distributed in the wind region.
Both these assumptions are questionable.  We would, in principle, expect 
the metals to be contained in small clumps, as part of 
clumpy winds or fast supernova ejecta, possibly making mixing more efficient.  
In addition, the wind material should vary in composition from the supernova material, although
this would still mostly end up in the diffuse rather than the dense cold phase.
These are all complications that should be taken into account in future work. 

\acknowledgments
The numerical simulations were performed on the local SGIAltix 3700 Bx2, which was partly funded by the Cluster of
Excellence: “Origin and Structure of the Universe.”



\bibliographystyle{natbib}

\bibliography{ntor1111_metals}


\end{document}